\documentclass[lnicst]{svmultln}%
\usepackage{graphicx}
\usepackage{makeidx}  
\begin{document}
\mainmatter              
\title{A Framework for Designing\\3D Virtual Environments}
\titlerunning{Designing 3D Virtual Environments}  
%
\author{Salvatore Catanese\inst{1} \and Emilio Ferrara\inst{2} \and Giacomo Fiumara\inst{1} \and \\Francesco Pagano\inst{3}}
\authorrunning{S. Catanese, E. Ferrara, G. Fiumara, F. Pagano}   
%
\tocauthor{Salvatore Catanese, Emilio Ferrara, Giacomo Fiumara, Francesco Pagano}
\institute{	Dept. of Physics, Informatics Section, University of Messina, Italy 
						\and 
						Dept. of Mathematics, University of Messina, Italy 
						\and 
						Dept. of Information Technology, University of Milan, Italy}

\maketitle              


	
	
	
	
	

\begin{abstract}        
The process of design and development of virtual environments can be supported by tools and frameworks, to save time in technical aspects and focusing on the content.
In this paper we present an academic framework which provides several levels of abstraction to ease this work. 
It includes state-of-the-art components we devised or integrated adopting open-source solutions in order to face specific problems.
Its architecture is modular and customizable, the code is open-source.

\keywords {Virtual Environments, Games}
\end{abstract}

\section{Introduction}
Commercial games reach production costs up to millions dollars.
During the process of development, teams spend a lot of time in prototyping of particular features, and, commonly, building technical frameworks to design the game on top of them. 
Designing a game from scratch requires a lot of time, fund investments, skills and resources.
This process could be shortened by adopting existing platforms.
In this paper we describe a fully developed framework, thought to ease the development of 3D virtual worlds, supporting state-of-the-art techniques.
It could be adopted if programmers and artists would prefer to focus on aspects of the design process (e.g., gameplay, game mechanics, etc.), instead of technical aspects.
In academia, exploiting an existing solution could ease researchers to carry out specific experimentations ignoring other aspects of the development of a platform.
In fact, the source code of our framework has been released open-source\footnote{http://informatica.unime.it/velab}.
This represents a great advantage w.r.t. using different solutions, also freely available on the Web, whose code is closed, architecture is fixed and components are not replaceable, differently from our framework.
Moreover, its functionality and potential have already been exploited: demonstrative virtual environments, designed using our platform, have already been presented to the scientific community \cite{disio2011b}.

\section{Related work}
Literature and academic interest in supporting videogames development is growing. 
There are some works sharing similarities with ours, e.g., Liu et al. \cite{liu2009design} developed a Virtual Reality platform, designing server and client applications, including a rendering engine based on \textsl{OGRE}, combined with \textsl{RakNet} for networking features, and finally they developed an example application. 
Also Braun et al. \cite{braun2010vhcve} recently presented a platform, for simulating virtual environments where users communicate and interact each other, using avatars with facial expressions and body motions. 
Some interesting techniques of computer vision and physics effects have been implemented to increase the realism of the simulation.
In a broader panorama, Hu-xiong et al. \cite{hu2005solution} discussed the design and implementation of a game platform capable of running online games, both from client and server perspective. 
Concluding, Graham and Roberts \cite{graham2006toward} analyzed the videogame development process from a qualitative perspective, trying to define attributes of 3D games and adopting interesting criteria to help achieving desired standards of quality during the design of academic and commercial products.

\section{Architecture of the Framework}
In this section we briefly illustrate the components usually integrated in those frameworks supporting the design of virtual environments.
An important feature typically provided is the capability of rendering and managing 3D scenes, as 3D environments are usually enriched by a realistic behavior of elements populating them, by reproducing physics and detecting collisions. 
Input/output management includes, among others, reproducing music and sound, handling devices like keyboard and mouse, pads, etc. 
Characters populating 3D virtual worlds could be driven by artificial intelligence (another aspect optionally supported), by other humans (requiring networking features for multi-player aspects), or both.
In this project we developed components, interfaces and plug-in for the integration of some existing open-source solutions. 
We devised a framework able of supporting teams in the process of design and development of virtual environments.
Its modular nature ensures the possibility of integrating additional components to tackle specific requirements.

\subsubsection{Rendering Engine}
Our framework integrates an open-source rendering engine, namely \textsl{OGRE}.
The \emph{rendering} is the process of generation of an image from a model. 
A \emph{model} is a description of a three-dimensional object defined by data structures containing geometry, textures, lighting, materials, etc. 
The rendering can be in real-time or not: the first approach is adopted for example in developing videogames, while the latter is typically used in those applications where the primary requirement is photorealism rather than performance (e.g., post-production effects in movies, medical image analysis, etc.).
Rendering algorithms usually try to simulate optic phenomena to reproduce 3D environments.
These techniques rely on rendering \emph{primitives} (e.g., triangles and polygons for three-dimensional scenes), and not pixels.
The process of transforming 3D scenes into 2D images is implemented by rendering pipelines, supported by graphics hardware.
A typical input for a graphics pipeline is a model of scene and its output represents a bi-dimensional raster. 
\textsl{OpenGL} and \textsl{Direct3D} are two examples of graphics pipeline implementations.
The interaction between graphics pipeline and hardware is possible via direct access to resources or graphics libraries. 
Usually, rendering engines exploit existing graphics libraries to ease the process of development.
Our platform supports both \textsl{Direct3D} and \textsl{OpenGL}.

\subsubsection{Scene Manager}
A \emph{scene manager} is included in our framework to manage scenes representing 3D environments.
This component organizes the objects on the scene and the relationships among them, in a hierarchical way.
A scene manager could be designed in different ways and its implementation could affect the overall performance.
Our scene manager adopts a ``scene graph'' whose nodes represent entities or objects on the scene and edges represent relations among them. 
Moreover, it manages \emph{bounding volume hierarchies} (BVHs), trees adopted to represent bounding volumes (e.g., spheres, bounding boxes, etc.) containing objects.
BVHs are also adopted for speeding up the collision detection among objects on the scene.
In order to increase performance, minimizing the number of rendered elements, our framework supports techniques of spatial partitioning, such as the \emph{Binary Space Partitioning} (BSP) \cite{Fuchs80onvisible} and \emph{octrees} \cite{meagher1982geometric}. 
Some areas are not always visible in the frustum (i.e., the region of space visible from the character point of view), thus the scene manager, exploiting spatial relations among nodes, disregards their rendering.

\subsubsection{Physics Engine}
A valid framework should allow the simulation of the physics on the reproduced virtual environment. 
Several physics engines have been developed during last years to improve the degree of realism of videogames. 
\textsl{Havok} and \textsl{PhysX} are two examples.
The latter is a robust solution; we integrated it inside our framework, developing an interface to introduce a good level of abstraction.

\subsubsection{Collision Detection}
The problem of detecting collisions among objects on the scene is complex and involves several aspects of the simulation of a virtual world.
\textsl{PhysX} provides advanced directives for detecting collisions of characters with obstacles, to simulate the impact of objects with other elements on the scene and so on. 
\textsl{PhysX} adopts bounding volumes to surround objects inside shapes and checks for interactions among bounding volumes; it additionally exploits the ragdoll model for collision detection of characters.
After detecting a collision, the physics engine simulates effects on involved objects. 

\subsubsection{Character Controller}
One important aspect of the gameplay in a videogame is the strategy of control of the character. 
Players interact with the virtual world using their avatars.
It is fundamental to reproduce a realistic behavior in order to avoid frustration during the game experience. 
Moreover, the character is central in the perception of the player, thus all the imperfections are extremely visible.
An important aspect to be considered is the shape adopted for detecting the collisions.
A simple choise is a ``capsule'', embedding the character; this because, 
i) the shape is smooth, so the character could run on irregular terrains without getting stuck; 
ii) its symmetry ensures the character could turn around itself without hitting obstacles; 
iii) no rough-edges could obstacle the character when going through narrow passages. 
Collateral effects are related to its simplicity.
More complex models, such as the ragdoll \cite{witkin1997physically}, have been proposed.
This technique produces procedural animations, instead of static ones.
A ragdoll model is a collection of constrained-rigid-bodies (usually representing bones and joints of characters) adopted as a skeletal animation system.
Animations are generated in real-time, without adopting predetermined ones, increasing the degree of realism. 
Moreover, the rigid-body system is subjected to rules of the physics engine; thus, the interaction between the character and the environment is more accurate.
Our framework supports both these two control strategies.

\section{Design and Characteristics of our Platform}
Our goal was to design a framework solution to support design and development of virtual environments, providing the following features: 3D rendering capabilities, physics simulation, collision management and character controller, I/O management, scene and camera management (i.e., loading and saving scenes representing virtual environments, with graphics and physics characteristics, designed with a specific application).

\subsection{Adopted Tools}
There are several open-source solutions which provide state-of-the-art features for many of the required tasks.
Thus, the most of the time has been spent for the integration of these components together, in order to obtain a unique framework.
All the components, interfaced each other, work as a substrate for developing an application on top of them.
A short description of the chosen tools follows.

\subsubsection{OGRE}
It is a cross-platform 3D rendering engine.
This engine is scene-oriented and allows the graphical representation of 3D virtual environments, providing state-of-the-art techniques for visual effects, texturing and lighting.
A key aspect is its modular nature. 
It is possible to extend the provided features by the inclusion of new components.
We exploited both these aspects to extend and improve the engine itself. 
Its architecture supports both the \textsl{Direct3D} and \textsl{OpenGL} graphics pipelines.
\textsl{OGRE} is constituted by a collection of classes and libraries. 
Three main classes (\emph{scene manager}, \emph{resources manager} and \emph{render}) do the most of the work. 
It is possible to inherit and extend the default scene manager to implement new or different features; for example the ``BSPSceneManager'' is optimized to represent in-door environments, while the ``TerrainSceneManager'' better works with out-door scenes; finally, the ``OctreeSceneManager'' is a good compromise for general purposes.

\subsubsection{OGREOggSound} \label{OGREOggSound}
We integrated this component in the framework; it is an audio library which acts as a wrapper for the \textsl{OpenAL} API. 
Its adoption allows to automatize the inclusion of audio features, e.g., to support wave and ``Ogg Vorbis'' audio sources, static and dynamic environmental sounds, 2D/3D audio, etc. 
Sound elements can be included inside the scenes as positional audio sources to reproduce more realistic environments.

\subsubsection{PhysX and NxOGRE}
\textsl{PhysX} is part of our framework. 
It is in charge of the simulation of physics laws to increase the degree of realism of the reproduced virtual environments.
Some key aspects managed by the engine involve a rigid-body and soft-body system simulation, an advanced character controller supporting different shapes (e.g., capsules, boxes, triangular and convex meshes, etc.), simulation of fluid dynamics, field strength, collisions, etc.
\textsl{PhysX} is integrated via the \textsl{NxOGRE} class wrappers, which introduces a useful level of abstraction to ease the access to functionalities provided by the library.

\subsubsection{Blender}
\textsl{Blender} is a cross-platform open-source 3D graphics and modeling application. 
We adopted \textsl{Blender} because of the possibility of customizing this tool to include new features.
\textsl{Blender} can manage multiple scenes.
Each scene is represented by its structure, objects, textures, materials, sound sources, etc. 
We extended it to support also physics; some physical properties can be defined for each object, in order to better reflect its behavior in the virtual world and the type of interaction between the character and objects.

\subsection{The Framework}
The basic idea is to use \textsl{Blender} to design the scene and then pass its output to the appropriate manager.
\textsl{Blender} originally deals just with graphics, but we extended and improved its open and powerful architecture.
Using our framework, it is possible to add to each component of the scene a set of user-defined attributes and their values. 
This way, designers can specify values describing the physical properties of characters and objects on the scene.
To represent this combination of graphical and physical attributes, we defined a novel XML file format via a DTD (Document Type Definition).
Thus, we devised a new \textsl{Blender} exporter plug-in. 
The next step was to import these scenes into the framework. 
This task is delegated to the loader module which analyzes the imported file.
It sends graphical information to \textsl{OGRE}, physical information to \textsl{PhysX} and sounds to \textsl{OGREOggSound}. 
\textsl{OGRE} is responsible to manage the scene and to collect user input from devices. 
Thus, input is passed to \textsl{PhysX}, which manages the movement of characters, determines possible collisions and sends back to \textsl{OGRE} the response, in order to update the graphical representation. 
Another task was the camera management. 
The final touch was sound management, to add soundtracks and dynamic sounds.

\subsubsection{ExDotScene}
The first step, adopting our framework, is to create the virtual environment, which could be designed by using \textsl{Blender}.
Graphics are stored as meshes, scenes are represented adopting the designed scene graph, and physics is additionally included.
The standard \textsl{DotScene} format does not contain meshes or textures, but just an XML description of elements on the scene.
Our framework supports: 
i) the creation of physical objects with a graphical representation (bodies) and without (actors); 
ii) static and dynamic objects instantiation; 
iii) serialization of multiple 3D scenes.
Actually, the last point was not natively supported by the built-in DotSceneInterface library.
We developed an improved version, namely ExDotSceneInterface, to allow loading and saving multiple scenes supporting the following elements: 
i) nodes of the scene graph; 
ii) graphical entities (i.e., meshes, textures and materials); 
iii) lighting (i.e., lights on the scene with properties like positioning, direction, brightness, etc.); 
iv) camera (i.e., source node and target node pointed by the camera, its positioning and orientation); 
v) scene attributes (e.g., in/out-door, type of shading, clipping distance, etc.); 
vi) representation of the physics of the scene (e.g., serialization of static and dynamic actors and bodies, models of physics, etc.).
According to requirements of the scene loader, which should support the physics, a DTD has been defined, namely ExDotScene, which extends the official \textsl{DotScene} DTD.
A body element contains a shape element and an ``actorParam'' element. 
The shape element must contain one and only one element chosen among a list comprising cube, capsule, sphere, convex and triangular meshes.
These elements are described by typical attributes like dimensions, etc.
Moreover, it is possible to specify additional parameters for the shapes, using a ``shapeParam'' tag.

\subsubsection{Blender Exporter Plug-in}
This component recursively analyzes each node of the graph scene, considering information relative to its position, orientation and scale.
Also specific elements, such as lights, cameras or meshes are taken into account during this process.
Moreover, the framework gives the possibility to export only the properties of objects that satisfy some specific conditions.
This way, designers can choose to export physical properties related to the entire scene, or not.
Acting on the ``Logic Properties'' panel of \textsl{Blender}, they can set the following values in order to define physical properties of each object: 
i) \emph{body}, if set to false the object is an actor; 
ii) \emph{shape}, suggests the type of shape to use for the object; 
iii) \emph{static}, sets if the physical object is static or dynamic; 
iv) \emph{mass}, sets the mass -only for dynamic objects-; 
v) \emph{skin}, suggests the value of the skin width that the physics engine will use during the simulation; 
vi) \emph{file}, determines the .nxs file to be adopted as a physical representation of the convex or triangular shapes.
At the end of the export phase, the scene is stored as an XML file, using the previously discussed \textsl{ExDotScene} DTD, which includes the physics.

\subsubsection{Scene Importer} \label{Scene-Importer}
The core of the scene importer with physics support is composed of the DotSceneProcessor class, plus a number of other node processor classes.
DotSceneProcessor contains a list of objects and allows several scenes to be loaded.
Formats different from the \textsl{DotScene} are supported, albeit overriding the correct methods for loading the scene.
Our extension supports the ExDotScene format.

\subsubsection{Character Controller} \label{Character-Controller}
This component, which was designed exploiting some key aspects of the physics engine, includes the following features: 
i) creation of the principal and secondary characters (i.e., creating and tracking these instances, managing the interaction among them and between characters and the environment); 
ii) update of the character status (i.e., managing movements and actions, at each frame, and synchronizing the graphical mesh of the character w.r.t. the actual status); 
iii) character auto-stepping (to avoid characters get stuck in minor terrain bumps); 
iv) character walkable parts (i.e., definition of those areas of the environment that are not accessible to characters, acting like boundaries); 
v) modifiable bounding volumes (to simulate the crouching or groveling of the character); 
vi) character callback (i.e., response to collisions).
The control system of the character is composed of three main classes: GameCharacter, GameCharacterController and GameCharacterHitReport.
By using \textsl{PhysX}, it is possible to divide actors by groups, so as to manage and set different behaviors during the collision detection and response, accordingly.

\subsubsection{Game Character Controller}
The GameCharacterController implements the Singleton pattern so to ensure that only one instance of the class may exist inside the application.
This class is interfaced with the \textsl{PhysX} library and deals with the instantiation and the management of the character.
During the render loop of \textsl{OGRE}, the GameFrameListener class invokes a function of the GameCharacterController in order to execute the ``simulate'' and ``render'' methods over all characters recorded in the controller.

\subsubsection{Game Character} 
The GameCharacter class represents the player inside the physical scene.
It is possible to set the associated graphical mesh and the scene node in order to synchronize the visual and the physical representation.
A specific method deals with setting a new movement direction: a vector representing the three-dimensional components of the velocity of the character is used.
At a given moment the position of the character depends not only on the user input, but also on the velocity.
During the simulation cycle, the ``simulate'' method is used to update the physical shape and the ``render'' method to update the visual representation of the character by setting the position of the mesh according to the position, expressed in global coordinates, of the physical shape.

\subsubsection{Character Hit Report}
As a consequence of a collision, \textsl{PhysX} generates a HitReport event.
The GameCharacterHitReport class allows the setting of customized callback actions as a consequence of the occurrence of a collision, determining the actor with which the collision occurred and the group the actor belongs to.
If the actor is a dynamic object and belongs to a group the character may interact with, the impulse that must be applied to the actor is calculated.

\subsubsection{Camera}
Once a 3D virtual scene is designed and imported inside the framework, developers define the way users, using their characters, explore this world, the so called \emph{camera system}.
In general, the core of the camera system is composed of two scene nodes acting as point of view and target point of the camera.
We suppose that the node connected to the camera always points to the target node.
This way, the movement of the target node produces the movement of the camera node.
Moreover, it is possible to move the camera around the target object directly moving the camera node, thus obtaining particular cameraworks.
This simple system allows three cameras, namely \emph{chasing}, \emph{fixed} and \emph{first-person}.
In the \emph{chasing camera}, the target node is associated to the point the character is looking at.
A more distant view node makes the character as appearing displaced w.r.t. the center of the scene.
This way, the character is still visible but we obtain the effect of a side view of the scene.
The \emph{fixed camera} is similar.
The target node is the point the character is observing, but the camera is fixed and cannot be moved.
This system is used in several videogames (e.g., third-person games).
In \emph{first-person camera}, the scene node of the character is the camera node.
The camera is independent from any other object on the scene, thus the scene coordinates are used to update the position of camera and target nodes.

\subsection{Modular structure}
\begin{figure}[!ht]
	\centering
	\includegraphics[width=.75\columnwidth]{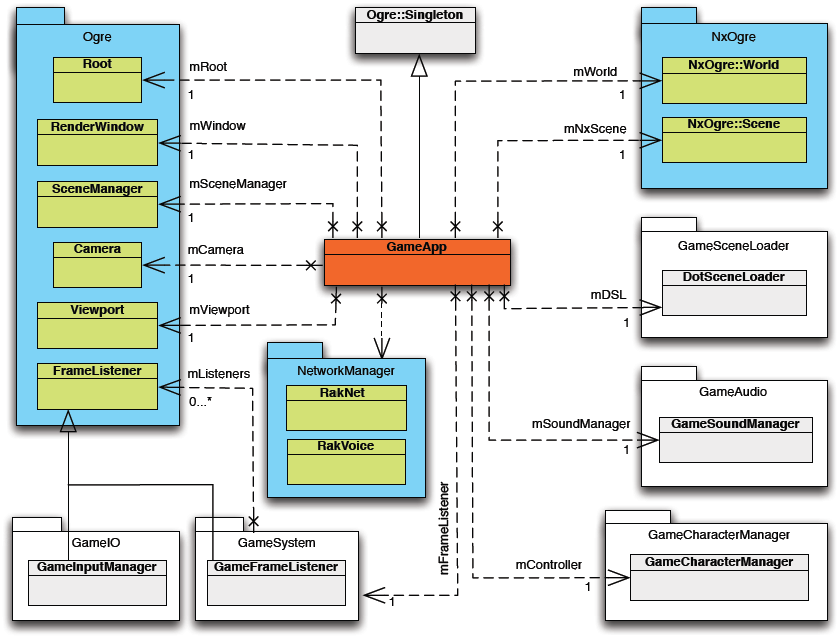}%
	\caption{The modular structure of the framework}%
	\label{fig:modular}%
\end{figure}

The framework has a modular structure, being composed of a series of packages each dealing with a certain functionality, as shown in Figure \ref{fig:modular}.
The binding process is obtained developing the required interface classes.
Sometimes, this introduces a useful level of abstraction in the implementation of the functionalities provided by each component.
Each package could be replaced with other components to satisfy particular requirements.
In other words, the framework can be customized without re-implementing all the functionalities, if the exposed interface of the replaced package is properly refactored.
This is one of the advantages while using our framework w.r.t. other common frameworks (e.g., Unity).
A brief description of the main packages follows.

\subsubsection{GameSystem}
It contains a series of classes dealing with the management of the rendering cycle.
Within \textsl{OGRE}, it is possible to define some classes which detect changes before and after a frame has been rendered on the screen.
To exploit this feature, it is necessary to register the various frame listeners in the object of \textsl{OGRE}.
The class GameFrameListener controls the order of execution of frame listeners, stored as an ordered list.
An instance of GameFrameListener is then registered in the root object, thus ensuring that each frame listener is executed in the correct order.

\subsubsection{GameIO}
It manages I/O devices using the Object-Oriented Input System (\textsl{OIS}) library.
\textsl{OIS} provides two modes to manage the input, namely unbuffered and buffered.
The latter input mode is safer, because the former could not reveal an event.
GameIO implements the buffered mode using the GameInputManager class.
GameKeyListener and GameMouseListener implement the \textsl{OGRE} corresponding interfaces 
and define the actions executed as a consequence of events produced by the input device(s).

\subsubsection{GameAudio}
It is developed as an interface with the \textsl{OpenAL} library in order to manage audio.
To do so, we developed a wrapper class using \textsl{OGREOggSound}, which provides methods to integrate \textsl{OpenAL} features (see Section \ref{OGREOggSound}).

\subsubsection{GameSceneLoader}
It provides those methods needed to load and save a scene.
It could be adopted to load \textsl{ExDotScene} based scenes, or, eventually, methods could be overloaded to manage different formats (see Section \ref{Scene-Importer}).

\subsubsection{GameCharacterController}
It is the interface with the control system of the character developed with \textsl{PhysX}.
It adopts a wrapper class to reproduce physics effects via \textsl{PhysX} and to manage the character controller (see Section \ref{Character-Controller}).

\subsubsection{RakNet}
We integrated a cross-platform library for networking, which provides support for TCP and UDP communications. 
It also includes \textsl{RakVoice}, a toolkit for VoIP support and for real-time communications during game sessions which relies on the sound engine to reproduce sounds.

\subsection{The Final Touch}
The virtual world we earlier described  still lacks of some features: first of all, logic has to be completely defined.
The framework does not substitute developers in this aspect; this because we believe that a valid product should be designed even in details and the game logic can not be a surrogate of preset patterns.
Moreover, the artificial intelligence of not-playing characters must be defined from scratch. 
The framework supports AI scripts coded in \textsl{Python}.
The framework provides default implementation for the management of I/O devices but is not necessarily the optimal solution for any requirement.
Similar arguments hold for other aspects, like the character control system or the camera system.
Concluding, the strength of our platform is its modular nature, which ensures that developers could take the best from each component, but could also replace functions using other components, providing better, or simply different functionalities.

\section{Conclusions}
In this paper we presented the design and implementation of an academic framework to support the development of 3D virtual environments, helpful in those cases in which teams would like to save time in technical aspects, to focus on the contents.
Our contribution can be summarized as follows: we presented a novel approach to create 3D virtual worlds enriched by physics effects. 
Our solution introduces the definition of physical properties of elements appearing on the scene, within the scene itself.
These directives are interpreted by the physics engine, which is strictly interfaced with the rendering engine, reproducing physics effects.
We additionally defined an extension of DTD for the \textsl{DotScene} format, introducing the support for the physics definition.
This work represents the basis for future extensions and has already been  adopted to show some techniques of design of virtual environments \cite{disio2011b}, e.g. the inclusion of environmental effects such as weather, day/night simulation and particle effects, exploiting techniques for terrain generation, realistic management of the water and fluid dynamics and the adoption of new rendering algorithms.
Future work includes its adoption in different fields of application such as i) virtual/augmented reality; ii) virtual tours, reconstructions and museums; iii) engineering and architectural simulations.

\bibliographystyle{splncs03}
\bibliography{sccg2011bib}
\end{document}